\documentclass[prd,nofootinbib,superscriptaddress]{revtex4}
\usepackage{amsmath}
\usepackage{breqn}
\usepackage{graphicx, subfig}
\usepackage{color}

\newcommand{\beq}{\begin{equation}}
\newcommand{\eeq}{\end{equation}}
\newcommand{\dd}{\partial}

\begin{document}

\title{Shapiro Delays at the Quadrupole Order for Tests of the
No-Hair Theorem Using Pulsars around Spinning Black Holes}
\author{Pierre Christian}
\affiliation{Harvard-Smithsonian Center for Astrophysics$,$ 60 Garden Street$,$ Cambridge$,$ MA$,$ USA}

\author{Dimitrios Psaltis}
\affiliation{University of Arizona$,$  933 N. Cherry Ave.$,$ Tucson$,$ AZ$,$ USA}

\author{Abraham Loeb}
\affiliation{Harvard-Smithsonian Center for Astrophysics$,$ 60 Garden Street$,$ Cambridge$,$ MA$,$ USA}

\date{Month date 2015}

\begin{abstract}
One avenue for testing the no-hair theorem is obtained through timing a pulsar
orbiting close to a black hole and fitting for quadrupolar effects on
the time-of-arrival of pulses. If deviations from the Kerr quadrupole
are measured, then the no-hair theorem is invalidated. To this end, we
derive an expression for the light travel time delay for a pulsar
orbiting in a black-hole spacetime described by the Butterworth-Ipser
metric, which has an arbitrary spin and quadrupole moment. We consider
terms up to the quadrupole order in the black-hole metric and derive
the time-delay expression in a closed analytic form. This allows for
fast computations that are useful in fitting time-of-arrival
observations of pulsars orbiting close to astrophysical black holes.
\end{abstract}

\maketitle

\section{Introduction}

The no-hair theorem of general relativity~\cite{NoHair} states that
the dimensionless quadrupole moment of a non-charged black hole, $q$, satisfies
\citep{tho80}
\beq
\label{eq:quadrupole_moment}
q \equiv \frac{c^4 Q}{G^2M^3} = - \left( \frac{c S}{G M^2}\right)^2 \;,
\eeq
where $M$, $S$, and $Q$ are the black-hole mass, spin angular
momentum, and quadrupole moment, respectively. If the quadrupole
moment of a black hole is measured to be in contention to
equation~(\ref{eq:quadrupole_moment}), one of the following
possibilities must occur: either the theory of general relativity
needs to be modified, or one of our assumptions regarding black-hole
solutions to the Einstein equation is invalid (e.g., the cosmic
censorship conjecture or the non-existence of closed timelike curves).

While the quadrupole $q$ has so far eluded measurement for any
astrophysical black hole, it may become accessible in the near future
with time-of-arrival (TOA) analysis of pulsars \citep{WK} orbiting close to the
supermassive black hole, Sgr~A*, at the center of the Milky
Way~\citep{pulsars,Liu}. The recent discovery of PSR~J1745$-$2900,
a magnetar orbiting close to Sgr~A*~\citep{sgr}, generated further
interest towards this possibility. While PSR~J1745$-$2900 is both too
unstable for precise time delay measurements~\citep{kaspi1} and
located too far from the black hole for relativistic effects to be
significant\footnote{Astrophysical implications of timing delays of
  pulsars at large distances from their black holes have been
  considered in a previous work \citep{chrisloeb}.}, the cluster of
stars around Sgr~A* should still harbor a significant number of
pulsars~\citep{pulsars,morepsr,P15}.

The effect of the quadrupole of a black hole on the orbit of a pulsar
around it has been studied in Refs.~\cite{Liu,WK,P15}. However,
calculations of higher-order effects on the propagation of light
itself and, in particular, on the Shapiro time delays in the pulsar
TOAs have been focused on lensing effects~\cite{LnR}. Similar
calculations of light time travel delays for solar-system experiments
have also been performed based on parametric post-Newtonian (PPN)
spacetimes with classical quadrupoles~\cite{RM, T2,ZnK}.

Testing the no-hair theorem of black holes (especially when dealing
with near-horizon tests) requires using special spacetimes that do not
have any pathologies near the horizon. In order to avoid pathologies,
such spacetimes do not have necessarily the same behavior as PPN
metrics at the quadrupole or higher order~\cite{metrics}.  The biggest
drawback in calculating the Shapiro time delay for a spacetime with an
arbitrary quadrupole moment is related to the fact that the presence
of a Carter-like constant is, in general, not guaranteed. Unlike the
case for Petrov-type~D spacetimes, such as the Kerr metric, the
Hamilton-Jacobi equation is not separable and the null geodesic motion
is challenging to solve. However, Ref.~\cite{TLPL} showed that the
coordinate travel time for a null geodesic obeys a Hamilton-Jacobi
like equation of motion that allows for the solution to be written in
terms of iterative integrals.

In this paper, we used this iterative approach to obtain an expression
describing the time delay of light as it propagates in the vicinity of
a black hole, taking into account the black-hole mass, spin, and
quadrupole moment. As a proof of principle, we use the Ricci flat
metric of Butterworth \& Ipser~\cite{BnI}, but the approach can be
easily generalized to any arbitrary metric with different far-field
expansions.  For this metric, we obtain an expression that is
analytical and allows for fast calculations to be performed. In \S2,
we describe our calculations; in the appendix, we compare our
calculations to previous results and, in \S3, we provide some
concluding remarks.

\section{Calculations}

\subsection{The metric and inverse metric to second order}

An asymptotically flat metric that is both stationary and axisymmetric
can be written up to order $(GM/rc^2)^2$ in the quasi-isotropic coordinates
$(t, r, \theta, \phi)$ as \citep{Morsink,BnI},
\begin{align} 
g_{tt} &= -1 + \frac{2 GM}{rc^2} - 2\left(\frac{GM}{rc^2}\right)^2 + {\cal O}\left[ \left( \frac{GM}{r c^2}\right)^3 \right]\;, \label{eq:metric1}
\\g_{rr} &= 1 + \frac{2 GM}{rc^2 } + \left( \frac{3}{2} - 2 \beta + 4 \beta \cos^2\theta \right) \left(\frac{GM}{rc^2}\right)^2 + {\cal O}\left[ \left( \frac{GM}{r c^2}\right)^3 \right]\;, \label{eq:metric2}
\\g_{\phi \phi} &= r^2 \sin^2\theta + r^2 \sin^2\theta \frac{2G M}{rc^2} + r^2 \sin^2\theta \left(\frac{3}{2} + 2\beta \right)\left(\frac{GM}{rc^2}\right)^2 + {\cal O}\left[ \left( \frac{GM}{r c^2}\right)^3 \right]\;, \label{eq:metric3}
\\g_{\phi t} &= -2 \frac{a_* GM^2}{r^3c^2}  r^2 \sin^2\theta \left(1 + \frac{GM}{rc^2} \right) + {\cal O}\left[ \left( \frac{GM}{r c^2}\right)^3 \right] \; , \label{eq:metric4}
\\g_{\theta \theta} &= r^2 g_{rr} \; 
\end{align}
where $a_*\equiv cS/(G M^2)$ and, following \cite{Morsink}, we have
defined 
\begin{equation}
  \beta \equiv (1/4) + \tilde{B}_0/M^2\;,
  \label{eq:beta}
\end{equation}
where $\tilde{B}_0$ is a multipole of Ref.~\citep{BnI}, as the
dimensionless parameter characterizing the black-hole.  We will now
convert this to Cartesian coordinates, set $G=c=1$, use geometric
units (so that distances and times are measured in units of $M$), and
write the metric order by order. To first order, we get
\begin{equation}
g^{(1)}_{tt} =g^{(1)}_{xx}=g^{(1)}_{yy}=g^{(1)}_{zz}
= \frac{2 }{ \sqrt{x^2+y^2+z^2}} \; ,
\end{equation}
where the contravariant metric to first order 
\beq
g^{\mu \nu}_{(1)} = \eta^{\mu \alpha} \eta^{\nu \beta} g_{\alpha \beta}^{(1)} \; ,
\eeq
gives identical components. Similarly, the second order metric $g^{(2)}_{\mu \nu}$ is
\begin{align}
g^{(2)}_{tt} &= -\frac{2 }{ \left(x^2+y^2+z^2\right)} \; ,
\\ g^{(2)}_{xt} &=\frac{2 a_* y}{ (x^2 + y^2 + z^2)^{3/2}}  \; ,
\\ g^{(2)}_{yt} &=  -\frac{2 a_* x}{(x^2 + y^2 + z^2)^{3/2}}  \; .
\\ g^{(2)}_{xy} &= -\frac{4 x y \beta}{(x^2+y^2+z^2)^2} \; , 
\\g^{(2)}_{xx} &=  \frac{\left[x^2 (3-4 \beta )+\left(y^2+z^2\right) (3+4 \beta )\right]}{2  \left(x^2+y^2+z^2\right)^2}  \; ,
\\g^{(2)}_{yy} &= \frac{ \left[y^2 (3-4 \beta )+\left(x^2+z^2\right) (3+4 \beta )\right]}{2 \left(x^2+y^2+z^2\right)^2} \;,
\\g^{(2)}_{zz} &= \frac{ \left[(x^2+y^2) (3-4 \beta)+z^2 (3+4 \beta )\right]}{2  \left(x^2+y^2+z^2\right)^2}\;.
\end{align}
and the contravariant metric to second order
\beq
g^{\mu \nu}_{(2)} = \eta^{\mu \alpha} \eta^{\nu \beta} g_{\alpha \beta}^{(2)} + \eta^{\mu \alpha} g^{(1)}_{\alpha \beta} g_{(1)}^{\beta \nu}  \; ,
\eeq
becomes
\begin{align}
g_{(2)}^{tt} &= -\frac{6  }{ \left(x^2+y^2+z^2\right)} \;, 
\\ g^{xt}_{(2)} &=-\frac{2 a_* y}{ (x^2 + y^2 + z^2)^{3/2}}  \; ,
\\ g^{yt}_{(2)} &=  \frac{2 a_* x}{(x^2 + y^2 + z^2)^{3/2}}  \; ,
\\g^{xy}_{(2)} &= -\frac{4 x y \beta}{(x^2+y^2+z^2)^2} \; , 
\\g_{(2)}^{xx} &= \frac{\left[x^2 (11-4 \beta )+\left(y^2+z^2\right) (11+4 \beta )\right]}{2  \left(x^2+y^2+z^2\right)^2} \;,
\\g_{(2)}^{yy} &= \frac{ \left[y^2 (11-4 \beta )+\left(x^2+y^2\right) (11+4 \beta )\right]}{2  \left(x^2+y^2+z^2\right)^2} \;,
\\g_{(2)}^{zz} &= \frac{ \left[\left(x^2+y^2\right) (11-4 \beta )+z^2 (11+4 \beta )\right]}{2  \left(x^2+y^2+z^2\right)^2} \;. 
\end{align}

The geodesic equation involves terms, via the Christoffel symbols,
that are of second order in the metric and its derivatives. As such,
the equation for the null geodesics to second order will, in
principle, involve terms that are proportional to $M$ and $M^2$
(describing the effects of mass), to $a_*$ and $a_*^2$ (describing
frame dragging), to $\beta$ (describing the effects of the
quadrupole), and cross terms proportional to $a_* M$. 

\subsection{Parametrization of the pulsar orbit} 

In this paper, we concentrate on the effect of the black-hole metric
on the light propagation and treat the pulsar orbit
parametrically. Relativistic effects on the orbit, calculated in
Refs.~\cite{WK,Liu}, can be added to our calculation to lowest order
by adding time dependences on the orbital parameters. To focus on the
effects of the time delays along geodesics, we also neglect phenomena
that arise from the velocity of the pulsar.

\begin{figure}[t]
\centerline{\includegraphics[scale=0.5]{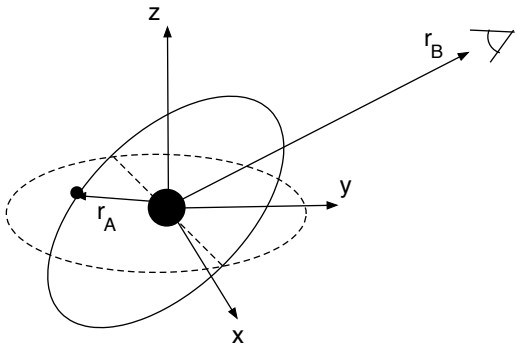}}
\caption{\label{fig:geometry} The geometry used in the
  calculation. The $z$-axis of the coordinate system is aligned with
  the spin of the black hole, the position vector of the pulsar is
  $\mathbf{r}_A$, and the position vector of the distant observer is
  $\mathbf{r}_B$ and lies on the $y-z$ plane.}
\end{figure}

In the following, we will set a Cartesian coordinate systems centered
on the black hole, with the $z-$axis parallel to the black-hole
angular momentum vector (see Figure~\ref{fig:geometry}). We also set
the $y$-axis such that the line connecting the black hole and the
observer lies on the $y-z$ plane (even though we write our expression
in a general vector notation that allows for an arbitrary orientation
of the observer). We focus our discussion on the light propagation
delay from the pulsar at position $\mathbf{r}_A$ to the observer at
$\mathbf{r}_B$.

For a pulsar in an eccentric orbit with semi-major axis $a$ and
eccentricity $e$, the magnitude of the distance between the pulsar and
black hole at an orbital phase corresponding to a true anomaly $\nu$
is given by \beq
\label{eq:pulsar_radius}
r_A = \frac{a (1-e^2)}{1 + e \cos\nu} \; ,
\eeq
while the direction of the vector $\mathbf{r}_A$ is given by
\begin{equation} \label{eq:lotsofrotations}
\hat{n}_A \equiv\frac{\mathbf{r}_A}{r_A}=  \mathbf{R_z}(\Omega) \mathbf{R_y}(i) \mathbf{R_z}(\nu) \mathbf{R_z}(\omega)  \mathbf{R_y}(-i)\cdot \begin{pmatrix}
0 \\
1 \\
0 
\end{pmatrix} \; ,
\end{equation}
where we have made use of the following definitions for the rotation matrices
\begin{equation}
\mathbf{R_y}(\theta) \equiv
 \begin{pmatrix}
  \cos \theta & 0 & \sin \theta \\
  0 & 1 & 0 \\
  -\sin \theta & 0 & \cos \theta 
 \end{pmatrix}   \;, \;\;\;\;
 \mathbf{R_z}(\theta) \equiv
 \begin{pmatrix}
  \cos \theta & -\sin \theta & 0 \\
  \sin \theta  & \cos \theta & 0 \\
  0 & 0 & 1 
 \end{pmatrix}   \; . 
\end{equation}
In this expression, $\omega$ is the argument of periapsis,
$\Omega$ is the longitude of the ascending node, and 
$i$ is the inclination of the orbit with respect to the black hole spin. 

Our expressions will depend on the angle between $\mathbf{r}_A$ and
$\mathbf{r}_B$, which we will leave expressed as the dot product
$\mathbf{n}_A \cdot \mathbf{n}_B$, where $\mathbf{n}_{B} \equiv
\mathbf{r}_{B}/r_{B}$. For most of the numerical examples shown in the
figures, we will set, for simplicity, the observer along the $y$ axis,
such that \beq \label{eq:nanb} \mathbf{n}_A \cdot \mathbf{n}_B =
\hat{y} \cdot\hat{n}_A =
\begin{pmatrix} 0 &1 &0
\end{pmatrix}\cdot \hat{n}_A \; .
\eeq
We further define the geometric distance between $\mathbf{r}_A$
and $\mathbf{r}_B$ as
\begin{align}
R_{AB} &\equiv \sqrt{r_A^2 +r_B^2 - 2 r_A r_B \mathbf{n}_A \cdot \mathbf{n}_B} \; .
\end{align}

\subsection{The first order Shapiro delay}

Because arbitrary stationary, axisymmetric spacetimes do not generally
admit a fourth constant of motion, solving analytically the null
geodesic equation of motion is difficult. However, it was recently
observed by Ref.~\cite{TLPL} that the coordinate time travelled by
light rays obeys Hamilton-Jacobi like equations that allows the light
propagation time delay to be written in terms of iterative
integrals. In particular, the propagation time delay to first order is
given by \citep{TLPL} \beq \Delta^{(1)} (\mathbf{r}_A, \mathbf{r}_B) =
\frac{1}{2} R_{AB} \int_0^1 \left[g_{(1)}^{00} - 2N^i_{AB}
  g_{(1)}^{0i} + N^i_{AB}N^j_{AB} g_{(1)}^{ij} \right]_{\mathbf{z}_+
  (\mu)} d \mu \;, \eeq where $N_{AB}^i = (r_B^i - r_A^i)/R_{AB}$ and
$\mathbf{z}_+ (\mu) = \mathbf{r}_A + \mu(\mathbf{r}_B -
\mathbf{r}_A)$.

Looking at the contravariant metric to first order, we can identify
the first and last term as the well known Shapiro delay effect for
non-rotating bodies (up to order 1). This is given by \citep{TLPL}
\begin{align} 
\Delta^{(1)}(\mathbf{r}_A, \mathbf{r}_B) &\equiv \int_0^1  \left[g_{(1)}^{00} + N^i_{AB}N^j_{AB}  g_{(1)}^{ij}  \right]_{\mathbf{z}_+ (\mu)}  d \mu \nonumber \\ 
&= 2  \log \left( \frac{r_A + r_B + R_{AB}}{r_A +r_B - R_{AB}}  \right) \;,
\label{eq:nospin1}
\end{align}
where $r_A$ and $r_B$ are the magnitudes of $\mathbf{r}_A$ and
$\mathbf{r}_B$ respectively.

In this section, we will express the magnitudes of the various effects
on the Shapiro delays in terms of their dependences on the Euclidian
distance of closest approach to the light ray from the black hole
\beq
r_c = \frac{r_A r_B}{R_{AB}} |\mathbf{n}_A \times \mathbf{n}_B | \;.
\eeq
We, therefore, rewrite equation (\ref{eq:nospin1}) as
\beq
\Delta^{(1)}(\mathbf{r}_A, \mathbf{r}_B) \approx 2 \log \left(
\frac{r_c+r_A \mathbf{n}_A \times \mathbf{n}_B}
     {r_c -r_A \mathbf{n}_A \times \mathbf{n}_B } \right) \; ,
\eeq
where we used the approximation that for astronomical applications,
$r_A << r_B$. This shows explicitly the known fact that the magnitude
of the first order Shapiro delay is logarithmic in $r_c$.

\subsection{The second order time delay}

The second order contribution to the light time travel delay is given
by~\cite{TLPL}

\beq \label{eq:TLPL}
\begin{split}
\Delta^{(2)} &(\mathbf{r}_A, \mathbf{r}_B) = \frac{1}{2} R_{AB} \int_0^1 \left\{ \left[g_{(2)}^{00} - 2N^i_{AB} g_{(2)}^{0i} +    N^i_{AB}N^j_{AB}  g_{(2)}^{ij}  \right]_{\mathbf{z}_+ (\mu)}  \right.
\\  & \left. + 2\left[N^j_{AB}g_{(1)}^{ij}   \right]_{\mathbf{z}_+ (\mu)} \frac{\dd \Delta^{(1)}}{\dd x^i}(\mathbf{x}_A, \mathbf{z}_+(\mu)) + \eta^{ij}\left[ \frac{\dd \Delta^{(1)}}{\dd x^i} \frac{\dd \Delta^{(1)}}{\dd x^j} \right]_{(\mathbf{x}_A, \mathbf{z}_+ (\mu))}   \right\}  d\mu \; .
\end{split}
\eeq This includes terms that are of second order in the mass and spin
of the black hole, as well as terms that are of first order in the
quadrupole.  As such, it describes the second order corrections to the
Shapiro time delays, the increase in the light paths because of
gravitational lensing, as well as the cross terms between these
effects. In this section we obtain analytical forms for this second
order time delay for spinning black holes with arbitrary quadrupoles.

\subsubsection{Mass contribution}

We will first consider the second order mass terms in equation
(\ref{eq:TLPL}). These are the terms in the second line of equation
(\ref{eq:TLPL}) together with the $g^{00}_{(2)}$ term in the first
line, which have been evaluated in Ref.\citep{TLPL}, i.e.,

\beq
\begin{aligned} \label{eq:Mass}
\Delta^{(2)}_{\rm mass} &= \frac{1}{2} R_{AB} \int_0^1 \left\{g^{00}_{(2)} +N^i_{AB}N^j_{AB}  g_{(2), M}^{ij}  +2\left[ N^j_{AB}g_{(1)}^{ij}   \right]_{\mathbf{z}_+ (\mu)} \frac{\dd \Delta^{(1)}}{\dd x^i}(\mathbf{x}_A, \mathbf{z}_+(\mu)) + \eta^{ij}\left[ \frac{\dd \Delta^{(1)}}{\dd x^i} \frac{\dd \Delta^{(1)}}{\dd x^j} \right]_{(\mathbf{x}_A, \mathbf{z}_+ (\mu))} \right\} d\mu  
\\&= \frac{1}{2} R_{AB} \left[\frac{15 \arccos(\mathbf{n}_A \cdot \mathbf{n}_B)}{2 r_A r_B \sqrt{1 - (\mathbf{n}_A \cdot \mathbf{n}_B })^2} - \frac{8}{r_A r_B (1 + \mathbf{n}_A \cdot \mathbf{n}_B) } \right]\; .
\end{aligned}
\eeq

Here, $g_{(2), M}^{ij} $ refers to the mass contribution to the
spatial metric, i.e., the terms that are not proportional to
the quadrupole parameter $\beta$. Equation (\ref{eq:Mass}) takes into
account the effect of gravitational lensing on the Shapiro
delay. Writing this second order mass contribution as
\beq
\Delta^{(2)}_{\rm mass} =  \frac{\vert\mathbf{n}_A \times \mathbf{n}_B\vert}{r_c}  \left[\frac{15 \arccos(\mathbf{n}_A \cdot \mathbf{n}_B)}{ 4 \sqrt{1 - (\mathbf{n}_A \cdot \mathbf{n}_B })^2} - \frac{4}{ (1 + \mathbf{n}_A \cdot \mathbf{n}_B) } \right] \; ,
\eeq
we find that this effect is of order $1/r_c$.

\begin{figure}[t]
  \centerline{\includegraphics[scale=0.45]{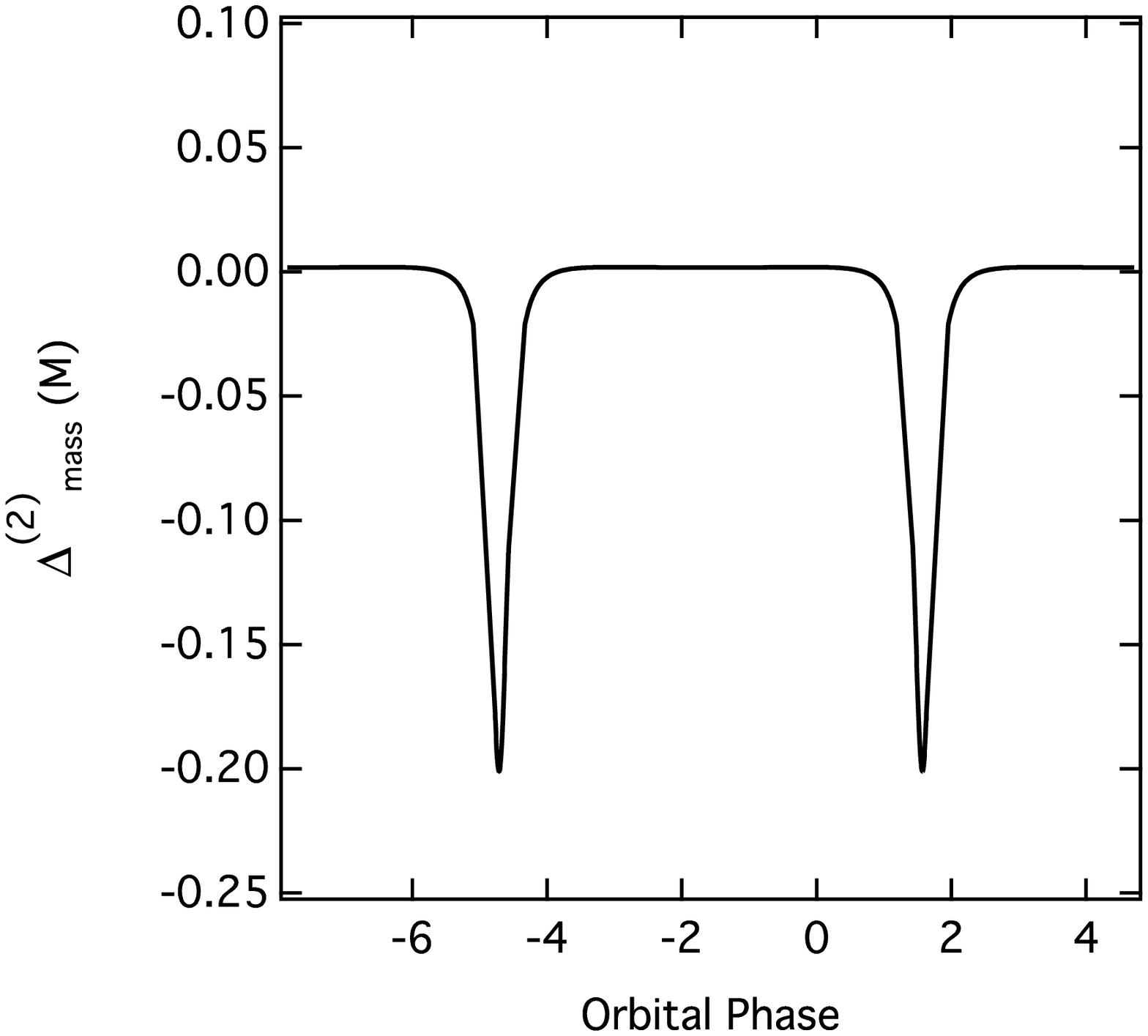}
  \includegraphics[scale=0.45]{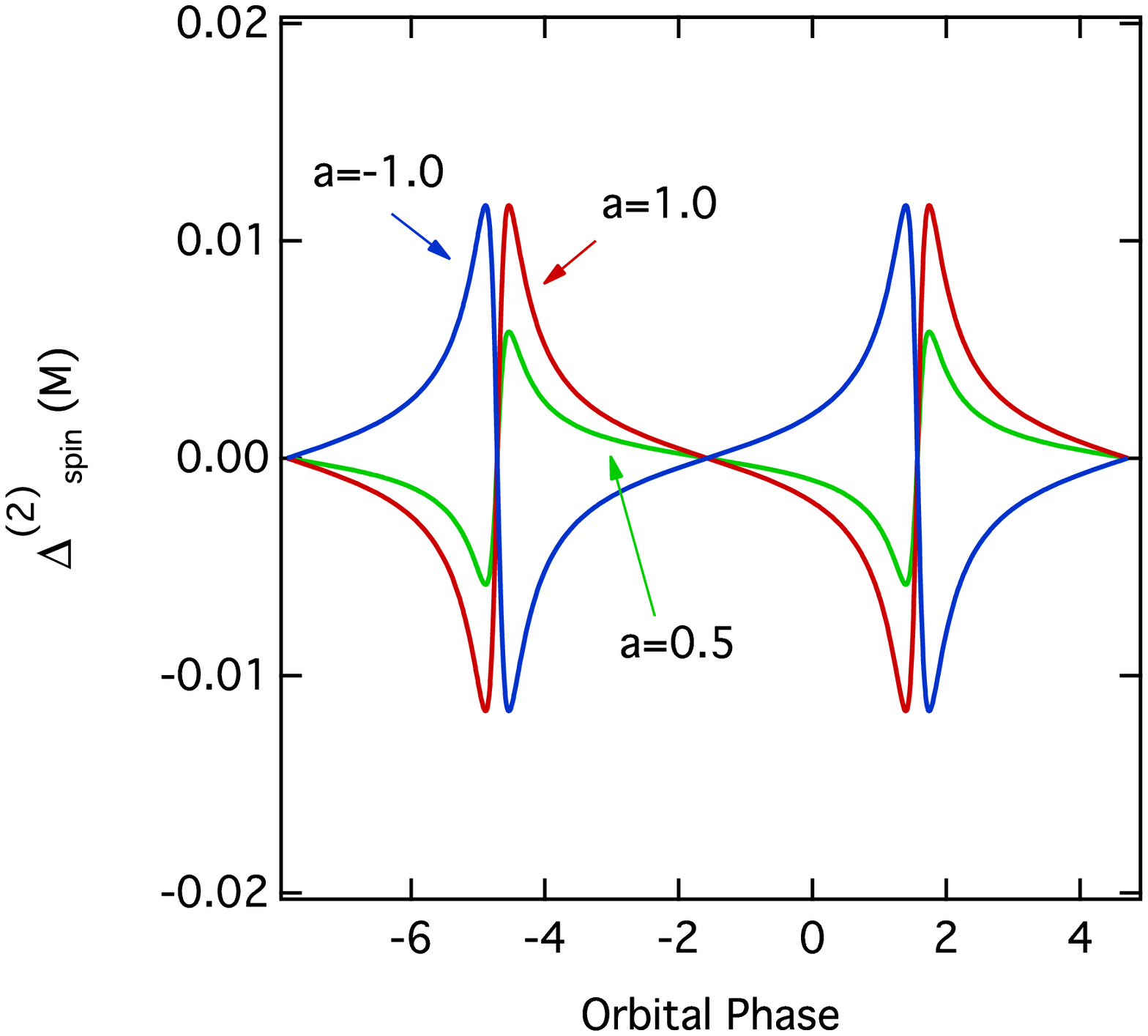}}
  \caption{\label{spin1order} The second-order contribution due to
    {\em (Left)\/} lensing and {\em (Right)\/} frame dragging to the
    light travel time delay for a pulsar in a circular orbit around a
    spinning black hole, as a function of orbital phase. The green and
    red lines in the right panel correspond to black-hole spins of
    $a=0.5$ and $a=1$, respectively, whereas the blue line corresponds
    to a black hole spinning at $a=1$ but in the opposite sense with
    respect to the pulsar orbit. In both panels, the pulsar orbital
    radius is $1000M$ and its inclination is $80$ degrees; the
    observer is set on the equatorial plane of the black hole;
    superior conjunction occurs at an orbital phase of $\pi/2$.}
\end{figure}

\subsubsection{Spin contribution}

The spin contribution to the second order light propagation delay is given by
\begin{align}
\Delta^{(2)}_{\rm spin} (\mathbf{r}_A, \mathbf{r}_B) &= \frac{1}{2} R_{AB} \int_0^1 \left[- 2N^i_{AB} g_{(2)}^{0i} \right]_{\mathbf{z}_+ (\mu)}  d \mu
\\&= -\int_0^1 \left[(x_B-x_A)\frac{2 a_* y}{(x^2 + y^2 +z^2)^{3/2}} - (y_B-y_A)\frac{2a_* x}{(x^2+y^2+z^2)^{3/2}}\right] d\mu \;.
\end{align}
Replacing $y$ and $x$ with $y = y_A + \mu(y_B-y_A)$ and
$x = x_A + \mu(x_B-x_A)$, we obtain 
\beq \label{eq:deltaspin1raw}
\Delta^{(2)}_{\rm spin} (\mathbf{r}_A, \mathbf{r}_B) = -\int_0^1 a_* \left[ \frac{-2 x_B y_A + 2 x_A y_B}{(x^2+y^2+z^2)^{3/2}}\right] d \mu \; .
\eeq

In order to perform this integral, we follow the integration scheme of
Ref.\cite{T2} with a small modification. We rotate the coordinate axis
to the plane defined by $\mathbf{r}_A$ and $\mathbf{r}_B$.
\beq
|\mathbf{z}_+(\mu)|=\sqrt{(x^2 + y^2 + z^2)}=\frac{r_c}{\cos(\gamma-\gamma_c)} ,
\eeq
where $\gamma$ is the angle between $\mathbf{r}_A$ and $\mathbf{r}_B$
(defined to be $0$ at $\mathbf{r}_B$) and $\gamma_c$ is the angle to
the point of closest approach. With these expressions and noting
that the differential can be expressed as
\beq
d\mu = \frac{|\mathbf{z}_+(\mu)|^2}{r_c R_{AB}} d\gamma \; ,
\eeq
the integral becomes
\begin{equation}
\int_0^1\frac{d\mu}{(x^2+y^2+z^2)^{3/2}}=
  \int_0^1 \frac{d\mu}{|\mathbf{z}_+|^{3}} = \int_{\gamma_A}^{\gamma_B} \frac{\cos{(\gamma-\gamma_c)}}{r_c^2 R_{AB}} d\gamma = \frac{r_A +r_B}{r_c^2 R_{AB}^2} (1 - \mathbf{n}_A \cdot \mathbf{n}_B) \; .
\end{equation}

Incorporating this to equation~(\ref{eq:deltaspin1raw}), we obtain
the second order spin correction to the propagation time delay,
\beq \label{eq:deltaspin1}
\Delta^{(2)}_{\rm spin} (\mathbf{r}_A, \mathbf{r}_B) = 2a_*(x_B y_A - x_A y_B) \left[ \frac{r_A +r_B}{r_A^2 r_B^2 } \frac{(1 - \mathbf{n}_A \cdot \mathbf{n}_B)}{|\mathbf{n}_A \times \mathbf{n}_B |^2}\right] \; .
\eeq
We can obtain the magnitude of the effect described by this equation by noting that
\beq
(x_B y_A - x_A y_B) = r_A r_B [\mathbf{n}_B - (\mathbf{n}_B \cdot \hat{z}) \hat{z} ] \times  [\mathbf{n}_A - (\mathbf{n}_A \cdot \hat{z}) \hat{z} ] \; , 
\eeq
allowing us to rewrite equation (\ref{eq:deltaspin1}) as 
\beq
\Delta^{(2)}_{\rm spin} (\mathbf{r}_A, \mathbf{r}_B) \approx \frac{2 a_*}{r_c}   \frac{[\mathbf{n}_B - (\mathbf{n}_B \cdot \hat{z}) \hat{z} ] \times  [\mathbf{n}_A - (\mathbf{n}_A \cdot \hat{z}) \hat{z} ] (1 - \mathbf{n}_A \cdot \mathbf{n}_B) }{\mathbf{n}_A \times \mathbf{n}_B}  \; .
\eeq
This demonstrates that the second-order effect on the Shapiro delay that is
due to frame dragging is of order $a_*/r_c$.

The right panel of Figure~\ref{spin1order} shows the second-order
contribution to the light time travel delay due to frame dragging for
a pulsar in a circular orbit around a black hole as a function of
orbital phase. We also compare it (left panel) to the second-order
contribution due to lensing derived in the previous section.  For the
purposes of this figure, we set the pulsar orbital radius to $1000M$,
its inclination to $80^\circ$, and the observer on the equatorial
plane of the black hole. We also varied the spin of the black hole
from being retrograde to the orbital motion ($a=-1$) to being prograde
($a=0, 0.4, 1.04)$.

As expected, the contribution due to frame dragging changes sign
around orbital phases $\nu=\pi/2$, as the photons from the pulsar to
the distant observer change from moving with the direction of frame
dragging to moving against it. The fact that at, these two phases in a
circular orbit, the contribution due to frame dragging vanishes while
the contribution due to lensing has its maximum value, makes the
overall amplitude of the former to be significantly suppressed
compared to the amplitude of the latter effect, even though they both
have the same scaling with $r_c$.

\subsubsection{Quadrupole contribution}

The quadrupole terms in equation (\ref{eq:TLPL}) are the
$g^{ij}_{(2)}$ terms that are proportional to the quadrupole
parameter, $\beta$. In order to evaluate them, we write
\beq \label{eq:quadRaw}
\begin{aligned}
\frac{1}{2}R_{AB} &\left[N^i_{AB}N^j_{AB}g^{ij}_{(2), \rm Q}\right]_{\mathbf{z}_+(\mu)}=
\\&\beta\frac{(x_B -x_A)^2}{R_{AB}}\frac{-x^2 +y^2+z^2}{(x^2 + y^2 +z^2)^2}
+ \beta\frac{(y_B-y_A)^2}{R_{AB}}\frac{-y^2 +x^2 +z^2}{(x^2 + y^2 +z^2)^2}
\\&-\beta \frac{(z_B-z_A)^2}{R_{AB}} \frac{-z^2 + x^2 +y^2 }{(x^2 + y^2 +z^2)^2}
-8\beta \frac{(x_B - x_A)(y_B - y_A)}{R_{AB}}\frac{x y}{ (x^2 + y^2 +z^2)^2}  \; ,
\end{aligned}
\eeq
where $g^{ij}_{(2), \rm Q}$ refers to the spatial metric components that are proportional to $\beta$. After substituting $\mathbf{r} = \mathbf{r}_A + \mu
(\mathbf{r}_B-\mathbf{r}_A)$, we perform the same mathematical trick as before,
but this time separating the terms proportional to $\mu^0$, $\mu$, and
$\mu^2$, and writing them in terms of their coefficients, $A$, $B$, and
$C$, i.e.,
\beq
\begin{aligned}
\frac{1}{2}R_{AB} \int_0^1&\left[N^i_{AB}N^j_{AB}g^{ij}_{(2)}\right]_{\mathbf{z}_+(\mu)} d\mu 
\\&= \int_0^1 \left[ \frac{A}{(x^2 + y^2 +z^2)^2 } + \mu \frac{B}{(x^2 + y^2 +z^2)^2} + \mu^2 \frac{C}{(x^2 + y^2 +z^2)^2 } \right] d\mu\; ,
\end{aligned}
\eeq
where
\beq
\begin{aligned}
  A&\equiv 2 [-4 x_A (x_A - x_B) y_A (y_A - y_B) - (x_A - x_B)^2 (x_A^2 - y_A^2 - 
      z_a^2) 
      \\&\;\;\;\;\;\;\;\;\;\;\;\;\;\;\;\;\;\;\;\;\;\;\;\;\;\;\;\;\;\;\;\;\;\;\;\;\;\;\;\;\;\;\;\; + (y_A - y_B)^2 (x_A^2 - y_A^2 + z_a^2) - (x_A^2 + y_A^2 - 
      z_a^2) (z_a - z_B)^2] \beta \\ 
  B&\equiv    4 \{x_A^4 - 3 x_A^3 x_B - 
   x_A x_B (x_B^2 + 3 y_A^2 - 4 y_A y_B + y_B^2 - z_a^2 + z_B^2) + 
   x_A^2 (3 x_B^2 + 2 y_A^2 - 3 y_A y_B + y_B^2 - z_a z_B + z_B^2) 
    \\&\;\;\;\;\;\;\;\;\;\;\;\;\;\;\;\;\;\;\;\;\;\;\;\;\;\;\;\;\;\;\;\;\;\;\;\;\;\;\;\;\;\;\;\; + (x_B^2 + 
      y_A^2 - 2 y_A y_B + y_B^2 + z_a^2 - 2 z_a z_B + z_B^2) [y_A^2 - y_A y_B + 
      z_a (-z_a + z_B)]\} \beta  \\
  C&\equiv 2 \{x_A^4 - 4 x_A^3 x_B + x_B^4 + y_A^4 - 4 x_A x_B [x_B^2 + (y_A - y_B)^2] + 
   2 x_A^2 [3 x_B^2 + (y_A - y_B)^2] + 2 x_B^2 (y_A - y_B)^2 
   \\&\;\;\;\;\;\;\;\;\;\;\;\;\;\;\;\;\;\;\;\;\;\;\;\;\;\;\;\;\;\;\;\;\;\;\;\;\;\;\;\;\;\;\;\; - 4 y_A^3 y_B + 
   6 y_A^2 y_B^2 - 4 y_A y_B^3 + y_B^4 - z_a^4 + 4 z_a^3 z_B - 6 z_a^2 z_B^2 + 
   4 z_a z_B^3 - z_B^4\}\beta \;.
\end{aligned}
\eeq

We can, therefore, write the second order contribution in this case as
\beq \label{eq:quadonly}
  \Delta^{(2)}_{\rm quad} (\mathbf{r}_A, \mathbf{r}_B) = \left[
    I_0 + I_1 + I_2 \right]_0^1\;,
\eeq
where the indefinite integrals
\begin{align}
I_0 &\equiv  \int \frac{A}{r^4} d\mu  \;, 
\\I_1 &\equiv  \int \frac{\mu B}{r^4} d\mu  \;, 
\\I_2 &\equiv  \int \frac{\mu^2 C}{r^4} d\mu \; ,
\label{eq:integr}
\end{align}
can be found in Appendix~A.

Figure~\ref{fig:quad_result} shows the second-order contribution due to
the quadrupole to the light travel time delay for a pulsar in a
circular orbit around a black hole. We consider black holes with
quadrupole moments characterized by $\beta =-0.5, \;0.2,$ and
0.5. The orbital distance of the pulsar is $1000M$, its inclination
is $80^\circ$, and the observer is placed on the equatorial plane of
the black hole.

The overall magnitude of the excursion due to quadrupole is much
smaller than lensing and frame-dragging contributions and increases
with the magnitude of the black-hole quadrupole. The complicated
expressions shown in Appendix~A make it hard to obtain the scaling of
this effect in an analytical manner. However, as we show in
Appendix~B, comparing our results with those of Ref.~\cite{ZnK}, which
were obtained using a different approach with harmonic coordinates,
allows us to simplify expression~(\ref{eq:quadonly}), for the
particular configuration that we are considering here as an example,
to
\begin{equation}
\Delta^{(2)}_{\rm quad} = - \frac{\beta}{2 R_{AB}} \left( \frac{r_A^{2} - r_B^{2} -
  R_{AB}^{2}}{r_B^{2}} + \frac{r_B^{2} - r_A^{2} -
  R_{AB}^{2}}{r_A^{2}} \right) \; .
\end{equation}
At the astrophysically relevant limit $r_B\gg r_A$, this expression reduces to
\begin{equation}
 \Delta^{(2)}_{\rm quad} =  - \frac{\beta}{r_A} \mathbf{n}_A \cdot \mathbf{n}_B \; .
  \end{equation}
Comparing this second-order contribution, which scale as the inverse
of the orbital distance to the pulsar, to the mass and spin effects
derived in the previous subsection, which scale as the inverse of the
distance of closest approach of light to the black hole, accounts for
the fact that the effect of the quadrupole is significantly smaller
for high-inclination observers than those of the mass and the spin.

\begin{figure}[t]
\centering
  \includegraphics[scale=0.45]{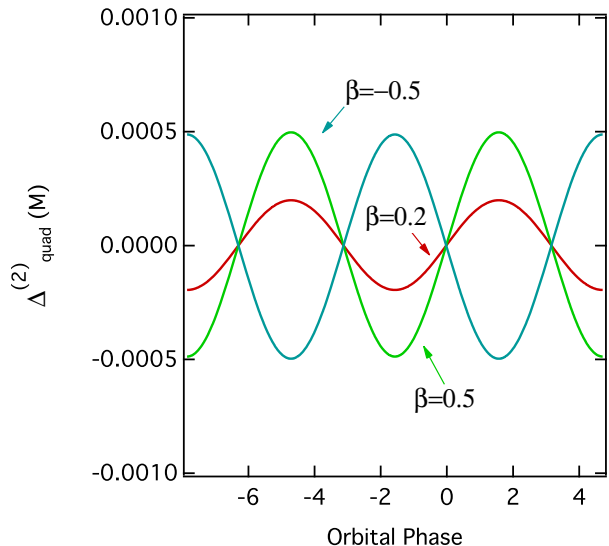}
\caption{\label{fig:quad_result} The second order quadrupole contribution
  to the light travel time delay for a pulsar in a circular orbit
  around black holes with different values of the quadrupole parameter
  $\beta$. The orbital radius of the orbit is $1000 M$, its
  inclination is $80^\circ$, and the observer is at the equatorial
  plane of the black hole;
    superior conjunction occurs at an orbital phase of $\pi/2$.}
\end{figure}

\section{Conclusion}

In this paper we have provided formulae for the light time travel
delays for pulsars orbiting in the spacetime of a black hole, taking
into account terms that are up to the quadrupole order.  We identified
three effects that are, in principle, of the same order.  The first
effect is expressed in terms that are proportional to the square of
the black-hole mass and describe the additional delays due to the
lensed trajectories of the photons. The second effect is expressed
in terms that are proportional to the black-hole spin and describe
the effects of frame dragging. Finally, the last effect describes
the influence of the mass quadrupole of the spacetime on the time
delays.

We reproduce our expression here for ease of reading, under the
astrophysically relevant assumption $r_B\gg r_A$:
\begin{eqnarray}
  \Delta^{(1)} &=&
  2 \log \left(
\frac{r_c+r_A \mathbf{n}_A \times \mathbf{n}_B}
     {r_c -r_A \mathbf{n}_A \times \mathbf{n}_B } \right) \; ,\nonumber\\
\Delta^{(2)}_{\rm mass} &=&\frac{\mathbf{n}_A \times \mathbf{n}_B}{r_c}  \left[\frac{15 \arccos(\mathbf{n}_A \cdot \mathbf{n}_B)}{ 4 \sqrt{1 - (\mathbf{n}_A \cdot \mathbf{n}_B })^2} - \frac{4}{ (1 + \mathbf{n}_A \cdot \mathbf{n}_B) } \right] \;,\nonumber\\
\Delta^{(2)}_{\rm spin}&=& \frac{2 a_*}{r_c}   \frac{[\mathbf{n}_B - (\mathbf{n}_B \cdot \hat{z}) \hat{z} ] \times  [\mathbf{n}_A - (\mathbf{n}_A \cdot \hat{z}) \hat{z} ] (1 - \mathbf{n}_A \cdot \mathbf{n}_B) }{\mathbf{n}_A \times \mathbf{n}_B}  \nonumber\\
\Delta^{(2)}_{\rm quad} &=&
\left[    I_0 + I_1 + I_2 \right]_0^1\;.
\end{eqnarray}

\begin{figure}[t]
\centering
  \includegraphics[scale=0.45]{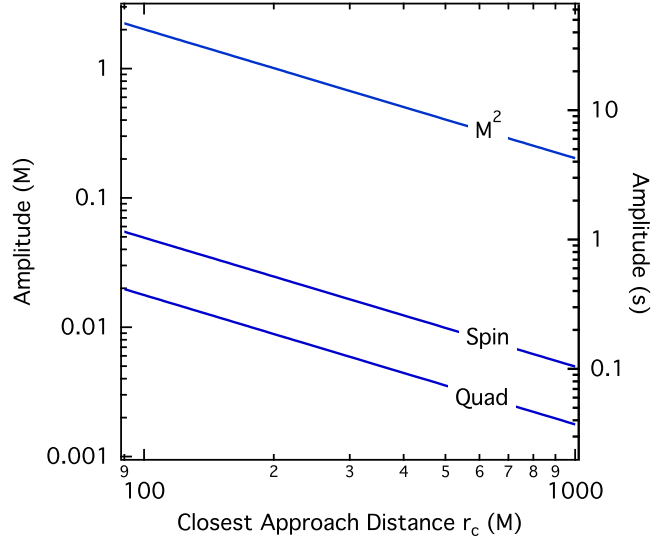}
  \caption{\label{fig:sgrA} The amplitudes of the various
    contributions to the light travel time delay for a pulsar in
    different circular orbits around a black hole, as a function of
    the closest approach distance $r_c$. The inclination of the orbit
    is $80^\circ$, the observer is at the equatorial plane of a Kerr
    black hole, and the spin of the black hole is maximal. The right
    axis shows the amplitudes of the various contributions in seconds,
    for the $4.3\times 10^6 M_\odot$ mass of Sgr~A*. Even though these
    higher-order effects are small compared to the traditional Shapiro
    delay, they are much larger than the expected measurement
    uncertainties for pulsars around Sgr~A*.}
\end{figure}

Even though the second-order effects are significantly smaller than
the traditional Shapiro delay, their amplitude for the case of a
pulsar orbiting a supermassive black hole is not negligible.  This is
shown in Figure~\ref{fig:sgrA}, where we plot the amplitude of each
effect (defined as the difference between the time delays calculated
at the points of superior and inferior conjunction) for different
pulsars orbiting the black-hole in the center of the Milky Way,
Sgr~A*. For reasonable distances of closest approach (see, e.g,
discussion in~\cite{pulsars, Liu, P15}), the amplitudes of these
effects are of the order of 100~ms$-$10~s. These are much larger than
the $\sim 1$~ms measurement uncertainties expected for observations of
a pulsar in orbit around Sgr A* with a 100-m dish or the $\lesssim
0.1$~ms uncertainties expected with SKA~\cite{Liu}.

At large distances from Sgr A*, the time delay in the pulsar signal may be
contaminated by the presence of additional mass between the pulsar and the black
hole. The ratio between the leading second-order terms in the metric
due to the gravitational field of the black hole and the first-order
terms due to the additional enclosed mass scale as \cite{P15}
\beq
\frac{M_{\rm enc}}{M}\left( \frac{ac^2}{GM} \right) =
  4.8 \times 10^{-8} \left( \frac{M_{\rm enc}}{10^6 \; M_\odot } \right)
  \left( \frac{a_0}{1 \; {\rm pc}} \right)^{-1} \left( \frac{a c^2}{G M} \right)^2 \; ,
\eeq
where $M_{\rm enc}$ is the enclosed mass within a distance $a_0$ from the black hole
and we have assumed a radial profile in the density of matter proportional to $r^{-2}$. 
In order for the gravitational effects of the enclosed mass to be negligible
compared to the second-order effects due to the gravitational field of
the black hole, the above ratio has to be smaller than unity, or
\beq
\frac{a c^2}{G M}\lesssim  4500
\left( \frac{M_{\rm enc}}{10^6 \; M_\odot } \right)^{-1}
  \left( \frac{a_0}{1 \; {\rm pc}} \right)\;.
\eeq
For pulsars in orbits with larger separations from the black hole,
the second order effects we calculated here will not be measurable.

The magnitude of the effects shown in Figure~\ref{fig:sgrA} are 
larger than the orbital effects due to the spacetime quadrupole that
were discussed in Ref.~\cite{P15}. Neglecting the high-order time
delay effects, however, will primarily introduce a small bias to the
measurement of the quadrupole discussed in Ref.~\cite{P15} depending
on the orientation of the pulsar orbit and observer, since the
quadrupole-order time-delay and orbital effects have very different
signatures on the pulsar TOAs.

We have also chosen to stop our expansion keeping terms up to the
quadrupole order in the back-hole spacetime. Even though this is a
reasonable expansion when discussing the metric elements of a slowly
spinning black hole, it does not necessarily reflect an appropriate
expansion scheme when calculating observables that depend non-linerly
on the metric elements. Indeed, Refs.~\cite{Expansion} have explored
even higher order corrections to the light time travel delays and
found that the expansion converges as long as (incorporating back the
mass of the black hole) $2M r_A/r_c^2\ll 1$. This is an important
  condition to check when applying our results when the geometric mean
  of the pulsar orbital radius and the horizon size become comparable
  to the distance of closest approach of light.

\begin{figure}[t]
\centering
\includegraphics[scale=0.45]{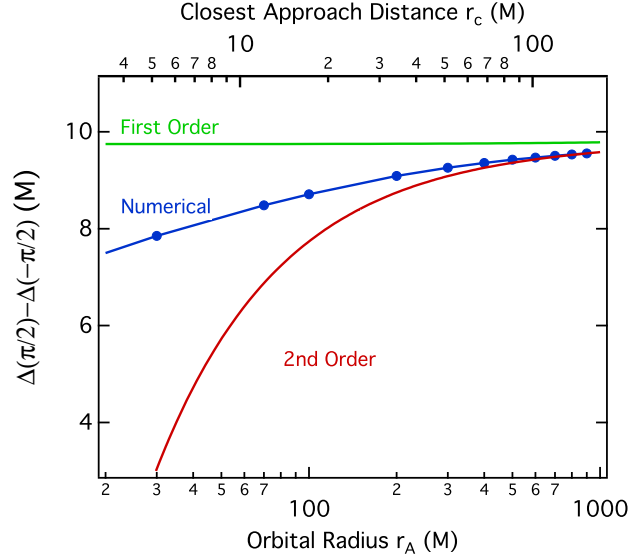}
\caption{\label{fig:deltaT} The difference between the light travel
  time delay at superior and inferior conjunction as a function of
  orbital radius for a pulsar in circular orbits around a non-spinning
  black hole at an inclination of $80^\circ$.  The green line is the
  first order Shapiro delay and the red line is our second order
  calculation. The blue line with the filled circles is the result of
  a numerical calculation using the algorithm of Ref.~\cite{PJ12}. The
  difference between the numerical result and the first order solution
  is significant even at large radii. The second order solution
  becomes inaccurate only at distances of closest approach that are
  $r_c\lesssim 60$~M.}
\end{figure}

As a final validity check of our analytic expansion, we compared our
analytic result to a numerical calculation of light time travel delays
using the numerical algorithm of Ref.~\cite{PJ12}. In
Figure~\ref{fig:deltaT} we plot the difference between the light
travel time at superior and inferior conjunctions as a function of
orbital radius, for a pulsar in circular orbits around a non-spinning
black hole at an inclination of $80$ degrees. It is clear even from
this comparison that second-order effects become important at orbital
radii that are of interest to pulsars around the black hole in the
center of the Milky Way. Moreover, this comparison demonstrates that
our second-order results remain accurate down to distances of closest
approach $r_c \sim 60 M$ and will, therefore, be useful in the analysis
of pulsar observations in this context.

\appendix

\section{Expression for the integrals of the quadrupole contribution}
The indefinite integrals required to calculate the propagation time delay to second order in equation (\ref{eq:integr}) is given by
\begin{multline*}
\int \frac{A}{r^4} d\mu =    \bigl(-4 x_A (x_A-x_B) y_A (y_A-y_B)-(x_A-x_B)^2 \bigl(x_A^2-y_A^2-z_A^2\bigr)\\ 
+(y_A-y_B)^2 \bigl(x_A^2-y_A^2+z_A^2\bigr)+\bigl(x_A^2+y_A^2+z_A^2\bigr) (z_A-z_B)^2\bigr) \beta  \bigl(\bigl(-z_A^2+z_A z_B+x_A^2 (-1+\mu )+y_A^2 (-1+\mu )+x_B^2 \mu +y_B^2 \mu +z_A^2 \mu \\
-2 z_A z_B \mu +z_B^2 \mu +x_A (x_B-2 x_B \mu )+y_A (y_B-2 y_B \mu )\bigr)/\bigl(\bigl(x_B^2 \bigl(y_A^2+z_A^2\bigr)+(y_B z_A-y_A z_B)^2-2 x_A x_B (y_A y_B+z_A z_B)+x_A^2 \bigl(y_B^2+z_B^2\bigr)\bigr)\\
 \bigl(z_A^2+x_A^2 (-1+\mu )^2+y_A^2 (-1+\mu )^2-2 z_A^2 \mu +2 z_A z_B \mu -2 x_A x_B (-1+\mu ) \mu -2 y_A y_B (-1+\mu ) \mu +x_B^2 \mu ^2+y_B^2 \mu ^2+z_A^2 \mu ^2-2 z_A z_B \mu ^2+z_B^2 \mu ^2\bigr)\bigr)\\
 +\left(\bigl(x_A^2-2 x_A x_B+x_B^2+y_A^2-2 y_A y_B+y_B^2+z_A^2-2 z_A z_B+z_B^2\bigr) \text{ArcTan}\left[\bigl(-z_A^2+z_A z_B+x_A^2 (-1+\mu )+y_A^2 (-1+\mu )s\right. \right.\\ 
 \left. \left. +x_B^2 \mu +y_B^2 \mu +z_A^2 \mu -2 z_A z_B \mu +z_B^2 \mu +x_A (x_B-2 x_B \mu )+y_A (y_B-2 y_B \mu )\bigr)/\bigl[(x_B^2 \bigl(y_A^2+z_A^2\bigr)+(y_B z_A-y_A z_B)^2 \right. \right. \\
\left. \left. -2 x_A x_B (y_A y_B+z_A z_B)+x_A^2 \bigl(y_B^2+z_B^2\bigr))^{(1/2)}\bigr]\right]\right)/\bigl(x_B^2 \bigl(y_A^2+z_A^2\bigr)+(y_B z_A-y_A z_B)^2-2 x_A x_B (y_A y_B+z_A z_B)+x_A^2 \bigl(y_B^2+z_B^2\bigr)\bigr)^{3/2}\bigr)
\end{multline*}

\begin{multline*}
\int \frac{\mu B}{r^4} d\mu =   2 \bigl(2 (x_A-x_B) (y_A-y_B) (2 x_A y_A-x_B y_A-x_A y_B)-\bigl(x_A^2-x_A x_B+y_A^2-y_A y_B+z_A (z_A-z_B)\bigr) (z_A-z_B)^2\\
+(y_A-y_B)^2 \bigl(-x_A^2+x_A x_B+y_A^2-y_A y_B+z_A (-z_A+z_B)\bigr)+(x_A-x_B)^2 \bigl(x_A^2-x_A x_B-y_A^2+y_A y_B+z_A (-z_A+z_B)\bigr)\bigr) \\
\beta  \bigl(\bigl(x_A^2 (-1+\mu )+y_A^2 (-1+\mu )-x_A x_B \mu -y_A y_B \mu +z_A (z_A (-1+\mu )-z_B \mu )\bigr)/\bigl(\bigl(x_B^2 \bigl(y_A^2+z_A^2\bigr)+(y_B z_A-y_A z_B)^2\\
-2 x_A x_B (y_A y_B+z_A z_B)+x_A^2 \bigl(y_B^2+z_B^2\bigr)\bigr) \bigl(z_A^2+x_A^2 (-1+\mu )^2+y_A^2 (-1+\mu )^2-2 z_A^2 \mu +2 z_A z_B \mu -2 x_A x_B (-1+\mu ) \mu -2 y_A y_B (-1+\mu ) \mu \\
+x_B^2 \mu ^2+y_B^2 \mu ^2+z_A^2 \mu ^2-2 z_A z_B \mu ^2+z_B^2 \mu ^2\bigr)\bigr)+ \left(\bigl(x_A^2-x_A x_B+y_A^2-y_A y_B+z_A (z_A-z_B)\bigr) \right. \\
\left. \times \text{ArcTan}\left[\bigl(-z_A^2+z_A z_B+x_A^2 (-1+\mu )+y_A^2 (-1+\mu )+x_B^2 \mu +y_B^2 \mu +z_A^2 \mu -2 z_A z_B \mu +z_B^2 \mu \right. \right. \\
\left. \left. +x_A (x_B-2 x_B \mu )+y_A (y_B-2 y_B \mu )\bigr)/\left(x_B^2 \bigl(y_A^2+z_A^2\bigr)+(y_B z_A-y_A z_B)^2-2 x_A x_B (y_A y_B+z_A z_B)+x_A^2 \bigl(y_B^2+z_B^2\bigr)\right)^{(1/2)}\right]\right)\\
/\bigl(x_B^2 \bigl(y_A^2+z_A^2\bigr)+(y_B z_A-y_A z_B)^2-2 x_A x_B (y_A y_B+z_A z_B)+x_A^2 \bigl(y_B^2+z_B^2\bigr)\bigr)^{3/2}\bigr)
\end{multline*}

\begin{multline*}
\int \frac{\mu^2 C}{r^4} d\mu =   -2 \bigl(x_A^4-4 x_A^3 x_B+x_B^4+y_A^4-4 x_A x_B \bigl(x_B^2+(y_A-y_B)^2\bigr)+2 x_A^2 \bigl(3 x_B^2+(y_A-y_B)^2\bigr)\\
+2 x_B^2 (y_A-y_B)^2-4 y_A^3 y_B+6 y_A^2 y_B^2-4 y_A y_B^3+y_B^4-z_A^4+4 z_A^3 z_B-6 z_A^2 z_B^2+4 z_A z_B^3-z_B^4\bigr) \\
\beta  \bigl(\bigl(-x_A^4+x_A^3 x_B-2 x_A^2 y_A^2+x_A x_B y_A^2-y_A^4+x_A^2 y_A y_B+y_A^3 y_B-2 x_A^2 z_A^2\\
+x_A x_B z_A^2-2 y_A^2 z_A^2+y_A y_B z_A^2-z_A^4+x_A^2 z_A z_B+y_A^2 z_A z_B+z_A^3 z_B+x_A^4 \mu -2 x_A^3 \\ x_B \mu +x_A^2 x_B^2 \mu +2 x_A^2 y_A^2 \mu -2 x_A x_B y_A^2 \mu -x_B^2 y_A^2 \mu +y_A^4 \mu 
-2 x_A^2 y_A y_B \mu +4 x_A x_B y_A y_B \mu -2 y_A^3 y_B \mu -x_A^2 y_B^2 \mu +y_A^2 y_B^2 \mu \\
+2 x_A^2 z_A^2 \mu -2 x_A x_B z_A^2 \mu -x_B^2 z_A^2 \mu +2 y_A^2 z_A^2 \mu -2 y_A y_B z_A^2 \mu \\
-y_B^2 z_A^2 \mu +z_A^4 \mu -2 x_A^2 z_A z_B \mu +4 x_A x_B z_A z_B \mu -2 y_A^2 z_A z_B \mu 
+4 y_A y_B z_A z_B \mu -2 z_A^3 z_B \mu -x_A^2 z_B^2 \mu -y_A^2 z_B^2 \mu +z_A^2 z_B^2 \mu \bigr)\\
/\bigl(2 \bigl(x_A^2-2 x_A x_B+x_B^2+y_A^2-2 y_A y_B+y_B^2+z_A^2-2 z_A z_B+z_B^2\bigr)\\
 \bigl(x_B^2 y_A^2-2 x_A x_B y_A y_B+x_A^2 y_B^2+x_B^2 z_A^2+y_B^2 z_A^2-2 x_A x_B z_A z_B-2 y_A y_B z_A z_B+x_A^2 z_B^2\\
 +y_A^2 z_B^2\bigr) \bigl(x_A^2+y_A^2+z_A^2-2 x_A^2 \mu +2 x_A x_B \mu -2 y_A^2 \mu +2 y_A y_B \mu -2 z_A^2 \mu +2 z_A z_B \mu +x_A^2 \mu ^2-2 x_A x_B \mu ^2\\
 +x_B^2 \mu ^2+y_A^2 \mu ^2-2 y_A y_B \mu ^2+y_B^2 \mu ^2+z_A^2 \mu ^2-2 z_A z_B \mu ^2+z_B^2 \mu ^2\bigr)\bigr)+\bigl(\bigl(x_A^2+y_A^2+z_A^2\bigr)\\
  \text{ArcTan}\bigl[\bigl(-x_A^2+x_A x_B-y_A^2+y_A y_B-z_A^2+z_A z_B+x_A^2 \mu -2 x_A x_B \mu +x_B^2 \mu +y_A^2 \mu -2 y_A y_B \mu +y_B^2 \mu +z_A^2 \mu -2 z_A z_B \mu +z_B^2 \mu \bigr)
  \\ /\bigl((\bigl(x_B^2 y_A^2-2 x_A x_B y_A y_B+x_A^2 y_B^2+x_B^2 z_A^2+y_B^2 z_A^2-2 x_A x_B z_A z_B-2 y_A y_B z_A z_B+x_A^2 z_B^2+y_A^2 z_B^2\bigr)\bigr) )^{1/2} \bigr]\bigr)\\
  /\bigl(2 \bigl(x_B^2 y_A^2-2 x_A x_B y_A y_B+x_A^2 y_B^2+x_B^2 z_A^2+y_B^2 z_A^2-2 x_A x_B z_A z_B-2 y_A y_B z_A z_B+x_A^2 z_B^2+y_A^2 z_B^2\bigr)^{3/2}\bigr)\bigr)
\end{multline*}

\section{Comparison with other calculations}

In this section, we compare the results of our calculations to those
of other analytic efforts that employed different approximations
and/or methods of solution. We also compare our results to numerical
calculations that take into account all multipole moments of a
spinning spacetime, in order to explore tha range of validity of our
approximations.

We do not attempt to compare our results to those of
Refs.~\cite{RnL,LnR} for two reasons. First, those calculations
combine the first-order Shapiro delay terms with the lensing equation,
making it hard to identify and compare the effects of individual
orders. Second, the lensing equation gives accurate results when the
pulsar is behind the black hole, at a distance that is much larger
than the distance of closest approach for light. This approximation is
valid only for a very narrow range of orbital phases (very close to
$\pi/2$) and observer inclinations (very close to $\pi/2$) and
introduces significant errors in more general configurations (see
discussion in~\cite{M11}).

Our results are in detail agreement with the PPN calculations of
Ref.~\cite{RM}, who also used a quasi-isotropic coordinate system,
when expressed in the appropriate variables.

Ref.~\citep{ZnK} calculated the light travel delay for the
Schwarzschild metric using the PPN formalism in de Donder (harmonic)
coordinates.  Their result is
\begin{align} \label{eq:PPN}
\Delta_{PPN} &= R^D_{AB} \nonumber
\\&+ 2 \log \left( \frac{r^D_A + r^D_B + R^D_{AB}}{r^D_A + r^D_B -R^D_{AB}} \right) \nonumber
\\ &+ 2 \frac{R^D_{AB}}{|\mathbf{r}^D_B \times \mathbf{r}^D_A|} \left[ (r^D_B - r^D_A)^2 - (R_{AB}^{D})^2 \right] \nonumber
\\ &+ \frac{15}{4} \frac{R^D_{AB}}{|\mathbf{r}^D_B \times \mathbf{r}^D_A |} \cos^{-1} (\mathbf{n}_A \cdot \mathbf{n}_B) \nonumber
\\ &+ \frac{1}{8} \frac{1}{R^D_{AB}} \left[ \frac{(r_A^{D})^2 - (r_B^{D})^2
  - (R_{AB}^{D})^2}{(r_B^{D})^2}  +  \frac{(r_B^{D})^2 - (r_A^{D})^2
  - (R_{AB}^{D})^2}{(r_A^{D})^2}  \right] \; ,
\end{align}
where the superscript $D$ denotes the r-coordinate in the de Donder
gauge.

The first term in the above expression corresponds to the geometric
delay, while the second term is the first order Shapiro delay.  This
second term is identical to our first order mass contribution to the
time delay given by equation (\ref{eq:nospin1}), even though they are
written in different coordinates. The reason is that the conversion
between the r-coordinates of de Donder, $r^D$, and the radial
Schwarzschild coordinate, $r_{\rm Sch}$, is (where for clarity, we
have temporarily reintroduced the black hole mass $M$ into our
equations)
\begin{equation}
  r^D = r_{\rm Sch} \left( 1 - \frac{M}{r_{\rm Sch}} \right)
  \; ,
\end{equation}
while the conversion between the isotropic radial coordinate (which we
use here), $r$, and the radial Schwarzschild coordinate, $r_{\rm Sch}$, is
\begin{align}
r &= \frac{r_{\rm Sch}}{2} \left[ 1 - \frac{M}{r_{\rm Sch}} + \left( 1 - 2
  \frac{M}{r_{\rm Sch}} \right)^{\frac{1}{2}} \right] \\& \approx r_{\rm Sch}
\left( 1 - \frac{M}{r_{\rm Sch} }\right) \; ,
\end{align}
As a result, to first order in $M/r$, $r \approx r^D$, and our expression
is algebraically identical to that of Ref.~\cite{ZnK}.

To second order in mass, the transformation between $r$ and $r^D$ is
given by
\begin{equation}
r \approx  r^D - \frac{M^2}{4 (r^{D})^2} \; .
\end{equation}
It is easily verifiable that plugging this transformations to our
equation for the second order mass term in the time delay does not
change the algebraic expression, i.e.,
\begin{equation} \label{eq:our_harm}
\Delta^{(2)}_{\rm mass}= \frac{R_{AB}^D}{2 r_A^D r_B^D}  \left[\frac{15 \arccos(\mathbf{n}_A \cdot \mathbf{n}_B)}{2 \sqrt{1 - (\mathbf{n}_A \cdot \mathbf{n}_B })^2} - \frac{8}{ (1 + \mathbf{n}_A \cdot \mathbf{n}_B) } \right] \; .
\end{equation}
Using the definition of $R_{AB}^D$, the third term of the delay in
equation (\ref{eq:PPN}) can be manipulated to read
\begin{equation}
2 \frac{R^D_{AB}}{|\mathbf{r}^D_B \times \mathbf{r}^D_A|^2} \left[ (r^D_B - r^D_A)^2 - R_{AB}^{D2} \right] =  -\frac{4 R_{AB}^D}{r_A^D r_B^D}  \frac{1}{1 + \cos \theta} \; ,
\end{equation}
which is the same as the second term of equation
(\ref{eq:our_harm}). Similarly, a trigonometric identity can be used
to transform the fourth term of the delay in equation~(\ref {eq:PPN})
into
\begin{equation}
\frac{15}{4} \frac{R^D_{AB}}{|\mathbf{r}^D_B \times \mathbf{r}^D_A |}
\cos^{-1} (\mathbf{n}_A \cdot \mathbf{n}_B) = \frac{15 R_{AB}^D}{4
  r_A^D r_B^D} \frac{1}{\sqrt{1 - \cos^2 \theta}} \; ,
\end{equation}
which is the same as the first term of equation (\ref{eq:our_harm}).

If the orbital configuration of the binary system is such that the $g_{zz}^{(2)}$ term can be ignored, i.e. when the term proportional to $(z_{B}-z_{A})^2$ in equation (\ref{eq:quadRaw}) is small, it is straightforward to identify the last term of equation (\ref{eq:PPN})
with the $\beta=-1/4$ case of equation (\ref{eq:quadonly}) and write
\beq
\Delta^{(2)}_{\rm quad} \left[I_0 + I_1 + I_2 \right]_{\beta = -1/4} = \frac{1}{8 R_{AB}} \left( \frac{r_A^{2} - r_B^{2} - R_{AB}^{2}}{r_B^{2}}  +  \frac{r_B^{2} - r_A^{2} - R_{AB}^{2}}{r_A^{2}}  \right) \; .
\eeq
We are inspired to seek a similar equivalence for arbitrary $\beta$, and by inspection we found that arbitrary quadrupole contributions of (\ref{eq:quadonly}) in this limit can be written as 
\beq \label{eq:identification}
\Delta^{(2)}_{\rm quad} =  - \frac{\beta}{2 R_{AB}} \left( \frac{r_A^{2} - r_B^{2} - R_{AB}^{2}}{r_B^{2}}  +  \frac{r_B^{2} - r_A^{2} - R_{AB}^{2}}{r_A^{2}}  \right) \; .
\eeq
This serves as a simplification of the complicated equation (\ref{eq:quadonly}), valid for all orbital configurations in the limit where the $g_{zz}^{(2)}$ term can be ignored. Indeed, for astrophysical purposes where $r_B \gg r_A$, the quadrupole delay in this limit is given by the extremely simple expression
\beq
\Delta^{(2)}_{\rm quad} |_{r_B \gg r_A} = - \frac{\beta}{r_A} \mathbf{n}_A \cdot \mathbf{n}_B \; .
\eeq

\acknowledgements

We thank Norbert Wex for many useful conversations on Shapiro delays
around black holes, for sharing with us M.\ Ali's Masters thesis,
and for pointing us to the appropriate GAIA technical reports. We also
acknowledge support from a joint Arizona-Harvard NSF award.

\end{document}